\documentclass[conference]{IEEEtran}
\IEEEoverridecommandlockouts
\usepackage{cite}
\usepackage{amsmath,amssymb,amsfonts}
\usepackage{algorithmic}
\usepackage{algorithm}
\usepackage{graphicx}
\usepackage{url}
\usepackage{booktabs} 
\usepackage{multirow}
\usepackage{placeins}  

\usepackage{textcomp}
\usepackage[table,xcdraw]{xcolor}
\def\BibTeX{{\rm B\kern-.05em{\sc i\kern-.025em b}\kern-.08em
    T\kern-.1667em\lower.7ex\hbox{E}\kern-.125emX}}
    
\begin{document}

\title{

An Efficient GPU-based Implementation for Noise Robust Sound Source Localization

}

\author{
    \IEEEauthorblockN{Zirui Lin\IEEEauthorrefmark{1}, Masayuki Takigahira\IEEEauthorrefmark{2}, Naoya Terakado\IEEEauthorrefmark{2}, Haris Gulzar\IEEEauthorrefmark{3}, \\Monikka Roslianna Busto\IEEEauthorrefmark{3}, Takeharu Eda\IEEEauthorrefmark{3}, Katsutoshi Itoyama\IEEEauthorrefmark{1}, Kazuhiro Nakadai\IEEEauthorrefmark{1} and Hideharu Amano\IEEEauthorrefmark{4}}

    \IEEEauthorblockA{\IEEEauthorrefmark{1}Dept. of Systems and Control Engineering, School of Engineering, Tokyo Institute of Technology, Tokyo, Japan\\
    Email: \{linzirui, itoyama, nakadai\}@ra.sc.e.titech.ac.jp}
    
    \IEEEauthorblockA{\IEEEauthorrefmark{2}Honda Research Institute Japan Co., Ltd., Saitama, Japan\\
    Email: \{m.takigahira, naoya.terakado\}@jp.honda-ri.com}
    
    \IEEEauthorblockA{\IEEEauthorrefmark{3}NTT Software Innovation Center, Tokyo, Japan\\
    Email: \{haris.gulzar.cf, monikkaroslianna.busto.px, takeharu.eda.bx\}@hco.ntt.co.jp}

    \IEEEauthorblockA{\IEEEauthorrefmark{4}Dept. of Information and Computer Science, Keio University, Kanagawa, Japan\\
    Email: hunga@am.ics.keio.ac.jp}
}

\maketitle

\begin{abstract}
Robot audition, encompassing Sound Source Localization (SSL), Sound Source Separation (SSS), and Automatic Speech Recognition (ASR), enables robots and smart devices to acquire auditory capabilities similar to human hearing. Despite their wide applicability, processing multi-channel audio signals from microphone arrays in SSL involves computationally intensive matrix operations, which can hinder efficient deployment on Central Processing Units (CPUs), particularly in embedded systems with limited CPU resources.
This paper introduces a GPU-based implementation of SSL for robot audition, utilizing the Generalized Singular Value Decomposition-based Multiple Signal Classification (GSVD-MUSIC), a noise-robust algorithm, within the HARK platform, an open-source software suite. For a 60-channel microphone array, the proposed implementation achieves significant performance improvements. On the Jetson AGX Orin, an embedded device powered by an NVIDIA GPU and ARM\textsuperscript{\textregistered}Cortex\textsuperscript{\texttrademark}-A78AE v8.2 64-bit CPUs, we observe speedups of 5648.7$\times$ for GSVD calculations and 10.7$\times$ for the SSL module, while speedups of 4245.1$\times$ for GSVD calculation and 17.3$\times$ for the entire SSL module on a server configured with an NVIDIA A100 GPU and AMD EPYC\textsuperscript{\texttrademark} 7352 CPUs, making real-time processing feasible for large-scale microphone arrays and providing ample capacity for real-time processing of potential subsequent machine learning or deep leraning tasks.



\end{abstract}


\section{Introduction}
Audition is a critical aspect of human inter-individual communication~\cite{b0}. Similarly, sound is essential for robots and smart devices, providing information for navigation and interaction within their surroundings~\cite{b6,x1,x2,x4,x7,x8}. 
To enable robots and smart devices with these auditory capabilities, robot audition technology has emerged~\cite{b9-RA}, which primarily realizes Sound Source Localization (SSL), Sound Source Separation (SSS), and Automatic Speech Recognition (ASR) with microphone arrays, 
endowing robots and smart devices with human-like auditory capabilities—recognizing sound sources, perceiving dynamic environmental changes, receiving voice commands, and facilitating effective human-machine interaction~\cite{a9lollmann,x5,x6}.

Robot audition technology has a wide range of applications across various scenarios~\cite{lin-chips}, including embedded systems such as in-home robot assistants~\cite{b8}, disaster relief robots or drones~\cite{b8, b9}, and monitoring sounds for wild animal behavior analysis~\cite{b8, b11, b12}.

The majority of robot audition system implementations predominantly rely on central processing units (CPUs)\cite{a3michaud, a6valin, a8yamamoto, a9lollmann, b8, c7cpu, c8cpu-sss, b39, b41Hoshiba}. However, the high time consumption and computational cost associated with sound source localization (SSL) and sound source separation (SSS)—which involve intensive matrix operations—pose significant challenges for practical applications\cite{c1guo-cost, c2ishi-cost, c4nouri-cost, c5, c6, b21ghdss}. The study~\cite{lin-chips} identified matrix operation functions as major bottlenecks in CPU-based implementations. While CPUs can achieve real-time processing for small-scale microphone arrays, such as eight-channel arrays, their limited processing power becomes a bottleneck as the array size increases. For instance, in drone audition~\cite{b9}, this limitation prevents achieving high-quality SSL and SSS when scaling up the array size in highly noisy environments. Additionally, SSL and SSS can serve as preprocessing steps to extract audio features for subsequent machine learning (ML) or deep learning (DL) inference or training~\cite{b8}. However, CPU-based implementations often hinder inference by not allocating sufficient processing time, which impedes real-time performance for ML/DL tasks.

To address these challenges, an efficient implementation is essential for real-time processing of large-scale microphone arrays. This paper presents a GPU-based implementation of SSL using Generalized Singular Value Decomposition-based Multiple Signal Classification (GSVD-MUSIC), a noise-robust algorithm, within HARK~\cite{b8, nakadai2010hark}, the leading open-source platform for robot audition. Our approach aims to enhance the applicability of robot audition in embedded scenarios with low signal-to-noise ratio (SNR) environments. For a 60-channel microphone array, on the Jetson AGX Orin—an embedded device powered by an NVIDIA GPU and ARM\textsuperscript{\textregistered}Cortex\textsuperscript{\texttrademark}-A78AE v8.2 64-bit CPUs—speedups of 5648.7$\times$ for the GSVD part and 10.7$\times$ for the SSL module are observed, while speedups of 4245.1$\times$ for GSVD calculation and 17.3$\times$ for the entire SSL module are obtained on a server with an NVIDIA A100 GPU and AMD EPYC\textsuperscript{\texttrademark} 7352 CPUs. Our work enables real-time processing for large-scale microphone arrays and reserve spaces for achieving real-time subsequent ML/DL tasks.

\vspace*{-2mm}

\section{Related Work}
\label{sec:2}

Several studies have explored GPU and Field-Programmable Gate Array (FPGA)-based implementations of SSL for robot audition, utilizing their parallel computing capabilities to accelerate processing~\cite{b33nguyen, b33-2silva, b33-5, b35belloch, b35-2belloch, b35-3belloch, b39, lin-chips}. FPGA-based solutions are often preferred in simpler scenarios, such as indoor service robots, where small microphone arrays are sufficient, and low latency and power consumption are critical~\cite{b33nguyen, b33-5, b15, b17lin, lin-chips}. However, FPGAs are highly specialized in terms of parallelism and resource configuration, leading to fewer computational resources for large-scale computations. Moreover, compared to GPUs of the same generation, FPGAs typically offer bandwidth capabilities that are several orders of magnitude lower~\cite{d1-FPGA-bandwidth}, resulting in significant latency when handling large-scale audio data.

On the other hand, GPUs are more commonly used in ML/DL tasks, which makes them a more optimal choice for implementing efficient SSL, especially for large-scale microphone arrays or when SSL is used as a preprocessing step for ML/DL. Previous work has developed GPU-based SSL implementations focusing on optimizations tailored to specific GPU architectures and microphone array configurations to achieve real-time processing~\cite{b35belloch, b35-2belloch, b35-3belloch, b39}. However, these solutions lack flexibility and are not easily accessible, limiting their broader applicability. To address this, a more generalizable SSL design was proposed~\cite{lin-chips}, supporting various GPUs and microphone arrays while maintaining effective performance.

Despite these advancements, the Standard Eigenvalue Decomposition (SEVD)-MUSIC-based SSL approach suffers from poor noise robustness. Research~\cite{GEVD,GSVD} shows that when using an eight-channel microphone array, the SSL accuracy of SEVD-MUSIC sharply declines as the Signal-to-Noise Ratio (SNR) approaches zero, limiting its effectiveness in low-SNR environments such as drone audition. To mitigate this issue, GSVD-MUSIC was introduced, offering superior noise robustness and maintaining a high SSL accuracy even at an SNR of -10 with an eight-channel array. However, a GPU-based implementation for GSVD-MUSIC is still lacking.

To enhance the applicability of robot audition in embedded and low-SNR environments, ensure real-time processing, and reserve more time for subsequent ML/DL tasks, a GPU-based SSL implementation utilizing GSVD-MUSIC is essential.

\section{Algorithms}
\label{sec:3}


\subsection{GSVD-MUSIC}
When the SSL module the robot audition platform HARK, through which we implement our GPU-based solution, selects the GSVD-MUSIC algorithm via hyperparameters, the SSL calculation proceeds as follows. The input audio signal, captured by an $M$-channel microphone array, is transformed into the frequency domain using the short-time Fourier transform (STFT) before the SSL module, represented as $\mathbf{X}(\omega, f) \in \mathbb{C}^M$. Here, $\omega$ denotes the frequency bin, and $f$ represents the frame index, as defined in the following equation:
\begin{equation} \label{eq1}
\mathbf{X}(\omega, f) = \left[ X_1(\omega, f), X_2(\omega, f), \ldots, X_M (\omega, f) \right]^T.
\end{equation}

At the beginning, $\mathbf{X}(\omega, f)$ is passed to {\sl AddCorrelation}, the first main sub-function of the SSL module, which computes the instantaneous correlation matrix $\mathbf{R}_{ins}(\omega, f) \in \mathbb{C}^{M \times M}$ for each frequency bin, as expressed by the following equation:
\begin{align}
\mathbf{R}_{ins}(\omega, f) &=  \mathbf{X}(\omega, f)\mathbf{X}^*(\omega, f), \label{eq2}
\end{align}
where $(\cdot)^H$ denotes the complex conjugate transpose operator. This matrix represents the correlation of signals across different channels of the microphone array at a single time frame.

Subsequently, $\mathbf{R}_{ins}(\omega, f)$ from all time frames within a preset time period $T$ are passed to the sub-function \textit{NormalizeCorrelation} for normalization of the correlation matrices. When considering past frames, the normalization is calculated as follows:
\begin{equation} \label{eq3}
\mathbf{R}(\omega, f) = \frac{1}{T} \sum_{i=-\text{T} + 1}^{0} \mathbf{R}_{\text{ins}}(\omega, f + i).
\end{equation}
$\mathbf{R}(\omega, f)$ is a positive semi-definite matrix that approximates the channel correlations within the microphone array.

The sub-function \textit{SVD} performs GSVD on \( \mathbf{R}(\omega, f) \), which separates the signal and noise subspaces. GSVD is given by:
\begin{align}
\mathbf{K}^{-1}(\omega, f) \mathbf{R}(\omega, f) &= \mathbf{E}(\omega, f) \mathbf{\Lambda}(\omega, f) \mathbf{E}_r(\omega, f), \label{eq4} \\
\mathbf{E}(\omega, f) &= \left[ \mathbf{e}_1(\omega, f), \cdots, \mathbf{e}_M(\omega, f) \right]^T, \nonumber \\
\mathbf{\Lambda}(\omega,f) &= \mathrm{diag}\left( \lambda_1(\omega,f), \ldots, \lambda_M(\omega,f) \right), \nonumber
\end{align}
where \( \mathbf{K}^{-1}(\omega, f)\in \mathbb{C}^{M \times M} \) represents the inverse of the preset noise correlation matrix ($\mathbf{K}(\omega, f)$), \( \mathbf{E}(\omega, f)\in \mathbb{C}^{M \times M} \) is the left singular vector matrix contains the left singular vectors ($\mathbf{e}_i(\omega, f)\in \mathbb{C}^{M}$),  $\mathbf{\Lambda}(\omega, f) \in \mathbb{C}^{M}$ is a diagonal matrix whose diagonal elements ($\lambda_i(\omega, f)$) are the singular values arranged in descending order and \( \mathbf{E}_r(\omega, f)\in \mathbb{C}^{M \times M} \) is the right singular vector matrix. Each singular value ($\lambda_i(\omega, f)$) quantifies the power contribution of its corresponding left singular vector ($\mathbf{e}_i(\omega, f)$). The subspace composed of left singular vectors with larger singular values is considered the signal subspace. For example, if there are \( N_s \) sound sources, the signal subspace is:
\[
\mathbf{E}_s = \left[ \mathbf{e}_1(\omega, f), \cdots, \mathbf{e}_{N_s}(\omega, f), \mathbf{0}, \cdots, \mathbf{0} \right], \mathbf{E}_s \in \mathbb{C}^{M \times M},
\]
and the noise subspace is:
\[
\mathbf{E}_n = \left[\mathbf{0}, \cdots, \mathbf{0}, \mathbf{e}_{N_s+1}(\omega, f), \cdots, \mathbf{e}_M(\omega, f) \right], \mathbf{E}_n \in \mathbb{C}^{M \times M}.
\]
The solution method for GSVD is detailed in Section~\ref{subsection:GSVD-Solution}.

\begin{figure}[H]
\centering
\includegraphics[scale=0.36]{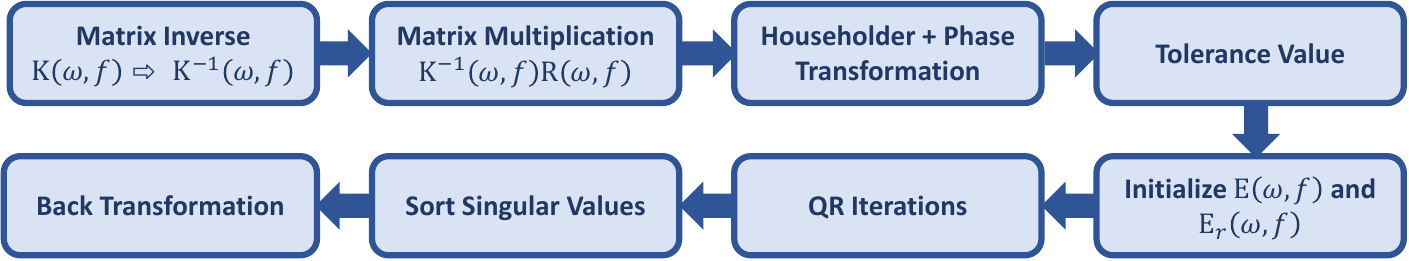}
\caption{Calculation Steps in the Sub-function \textit{SVD}.}
\label{fig:GSVD-Flow}
\end{figure}
\vspace*{-2mm}

Afterward, the sub-function \textit{CalcAveragePower} computes the power spectrum for each target direction using the MUSIC algorithm:
\begin{align} \label{eq5}
\mathbf{P}(\theta, \omega, f) &= \frac{\left| \mathbf{H}^H(\theta, \omega) \mathbf{H}(\theta, \omega) \right|}{\sum_{i=N_s+1}^{M} \left| \mathbf{H}^H(\theta, \omega) \mathbf{e}_n(\omega, f) \right|},
\end{align}
where  $\mathbf{H}(\theta, \omega) \in \mathbb{C}^M$ represents the transfer function between the microphone array and sound sources at the target directions, reflecting the array’s response to sounds from different directions and frequencies. When the target direction aligns with the actual sound source, the power spectrum exhibits a local maximum, as the audio signal is orthogonal to the noise subspace.

The power spectrum for each target direction ($\mathbf{P}(\theta, \omega, f)$) is then integrated in the frequency bin dimension as follows:
\begin{equation} \label{eq6}
\mathbf{\bar{P}}(\theta, f) = \sum_{\omega=\omega_{\min}}^{\omega_{\max}}  \mathbf{P}(\theta, \omega, f).
\end{equation}

Ultimately, the sound source directions are identified by searching for peaks in the directional range from  $\theta_{\text{min}}$ to $\theta_{\text{max}}$ of $\mathbf{\bar{P}}(\theta, f)$, based on the preset number of sound sources to be considered.

\subsection{GSVD Solution}
\label{subsection:GSVD-Solution}
In general, the solution methods for GSVD can be classified into two primary categories: 1) Jacobi iteration-based methods, which gradually diagonalize the matrix through a series of Givens rotations. These methods are well-suited for small-scale matrices but are inefficient for large-scale problems. 2) Methods based on Householder transformations and QR iterations, which are computationally efficient and better suited for large-scale matrices. Given that real-time processing for large-scale microphone arrays is a key objective, our implementation employs the latter approach. The main calculation steps involved in the solution process, within the sub-function \textit{SVD}, are outlined in Fig.~\ref{fig:GSVD-Flow}.

The process begins by calculating the inverse of the noise correlation matrix ($\mathbf{K}^{-1}(\omega, f)$) through Gaussian elimination. Then it computes the product of $\mathbf{K}^{-1}(\omega, f)$ and $\mathbf{R}(\omega, f)$, which is denoted as $\mathbf{A}(\omega, f)$. Subsequently, $\mathbf{A}(\omega, f)$ is diagonalized into a bidiagonal form using Householder transformations, and a phase transformation is also applied to ensure that the elements on both the main diagonal and subdiagonal are real and positive. The next step involves iterating over each column to compute the sum of the absolute values of the elements on the main diagonal and subdiagonal. A reduction operation is performed to identify the global maximum of these sums, which is then used to compute a tolerance value for the subsequent QR iterations, thereby determining the convergence of the matrix.
The left singular vector matrix ($\mathbf{E}(\omega, f)$) and the right singular vector matrix ($\mathbf{E}_r(\omega, f)$) are initialized as identity matrices. Then, QR iterations are performed until convergence, diagonalizing \( \mathbf{A}(\omega, f) \) and yielding the singular values. These singular values are then sorted in descending order, and the columns of \( \mathbf{E}(\omega, f) \) and \( \mathbf{E}_r(\omega, f) \) are reordered accordingly. Finally, a back transformation is applied to \( \mathbf{E}(\omega, f) \) and \( \mathbf{E}_r(\omega, f) \) to orthogonalize them after the column reordering.

\section{Methodology}
\label{sec:4}

\subsection{Development Environment}
To develop parallel computing for matrix operations involved in the SSL module, we utilized NVIDIA GPUs and CUDA\textsuperscript{\textregistered}. The parallelized matrix operation functions are transplanted to the SSL module of HARK to replace the original CPU-based functions.

\subsection{Implementation of Parallelization}
In HARK, the difference between SEVD-MUSIC-based SSL and GSVD-MUSIC-based SSL is the method to separate the signal and noise subspace, while other sub-functions are the same. Therefore, our work mainly focus on the parallelization of the GSVD part. For other matrix operations, we apply the implementation proposed in the work~\cite{lin-chips}, while the parallelization strategies for the GSVD part are illustrated in Algorithm~\ref{alg1}.

Lines 1-14 define and initialize the variables and functions. The calculation steps in Fig.~\ref{fig:GSVD-Flow} are represented by the functions \textit{MatInverse}, \textit{MatMul}, \textit{SVD}$_1$, \textit{SVD}$_2$, \textit{SVD}$_3$, \textit{SVD}$_4$, \textit{SVD}$_5$, and \textit{SVD}$_6$.

Lines 15-17 compute the inverse matrix \( \mathbf{K}^{-1} \) using Gaussian elimination. Since each row depends on the previous one, parallelism is limited to the frequency bin ($ \omega$) dimension.

Lines 18-20 compute the product \( \mathbf{K}^{-1} \mathbf{R} \) and assign it to \( \mathbf{A} \). As matrix multiplication is element-wise independent, parallelism is applied across the \( \omega \times M \times M \) dimension.

Lines 21-23 perform Householder and phase transformations to diagonalize \( \mathbf{A} \) into a bidiagonal form and ensure the elements are real and positive. While some loops in \textit{SVD}$_1$ could be further parallelized, we maintain the current parallelization in the \( \omega \) dimension to avoid synchronization overhead and memory access conflicts.

Lines 24-26 calculate the tolerance value (\( \epsilon \)). This step sums the absolute values of elements on the main diagonal and subdiagonal, which can be parallelized along the \( M \) dimension. A reduction operation is then used to find the global maximum. Hence, \textit{SVD}$_2$ is parallelized in the \( \omega \times M \) dimension, with a reduction.

Lines 27-29 initialize the left ($\mathbf{E}$) and the right ($\mathbf{E}_r$) singular vector matrix to identity matrices for each frequency bin component. This step is element-wise independent, the parallelism is applied across the \( \omega \times M \times M \) dimension, while \textit{SVD}$_2$ is responsible for the initialization of a single element.

\begin{algorithm}[H]
\caption{Parallel Computing of the sub-function \textit{SVD} using CUDA\textsuperscript{\textregistered}}
\label{alg1}
\begin{algorithmic}[1] 
\REQUIRE \textbf{Initialization:}
\STATE $\omega$ $\gets$ the considered frequency bin range
\STATE $M$ $\gets$ the number of microphones
\STATE $\mathbf{K}$ $\gets$ noise correlation matrix 
\STATE $\mathbf{R}$ $\gets$ audio correlation matrix 
\STATE $\mathbf{K}^{-1}$ $\gets$ audio correlation matrix 
\STATE $\epsilon$ $\gets$ tolerance value
\STATE \textit{MatInverse} function for the matrix inversion calculation
\STATE \textit{MatMul} function for calculating a single element in matrix multiplication
\STATE \textit{SVD}$_1$ function for Householder and phase transformation
\STATE \textit{SVD}$_2$ function for calculating the tolerance value ($\epsilon$)
\STATE \textit{SVD}$_3$ function for initializing a single element in the left ($\mathbf{E}$) and the right ($\mathbf{E}_r$) singular vector matrix
\STATE \textit{SVD}$_4$ function for QR iterations
\STATE \textit{SVD}$_5$ function for sorting singular values in descending order and adjusting the columns of $\mathbf{E}$ and $\mathbf{E}_r$
\STATE \textit{SVD}$_6$ function for the back transformation
\vspace{1mm}

\ENSURE The inverse matrix $\mathbf{K}^{-1}$.
\STATE dim3 block(32, 1)
\STATE dim3 grid(($\omega$-1)/block.x+1, 1)
\STATE \textit{MatInverse} \texttt{<<<}grid, block\texttt{>>>}($\mathbf{K}$)
\vspace{1mm}

\ENSURE $\mathbf{A}=\mathbf{K}^{-1}\mathbf{R}$
\STATE dim3 block(32, 4)
\STATE dim3 grid(($M$-1)/block.x+1, ($M$-1)/block.y+1, $\omega$)
\STATE \textit{MatMul}\texttt{<<<}grid, block\texttt{>>>}($\mathbf{K}^{-1}$, $\mathbf{R}$) 
\vspace{1mm}

\ENSURE $\mathbf{A}$ is transformed into bidiagonal form.
\STATE dim3 block(32, 1)
\STATE dim3 grid(($\omega$-1)/block.x+1, 1)
\STATE \textit{SVD}$_1$\texttt{<<<}grid, block\texttt{>>>}($\mathbf{A}$)
\vspace{1mm}

\ENSURE The tolerance value ($\epsilon$).
\STATE dim3 block(64, 1)
\STATE dim3 grid($\omega$, 1)
\STATE \textit{SVD}$_2$\texttt{<<<}grid, block\texttt{>>>}($\mathbf{A}$, $\epsilon$) 
\vspace{1mm}

\ENSURE $\mathbf{E}$ and $\mathbf{E}_r$ are identity initialized for each frequency bin.
\STATE dim3 block(32, 4)
\STATE dim3 grid(($M$-1)/block.x+1, ($M$-1)/block.y+1, $\omega$)
\STATE \textit{SVD}$_3$\texttt{<<<}grid, block\texttt{>>>}($\omega$, $M$)
\vspace{1mm}

\ENSURE $\mathbf{A}$, $\mathbf{E}$, and $\mathbf{E}_r$ are updated through QR iterations.
\STATE dim3 block(32, 1)
\STATE dim3 grid(($\omega$-1)/block.x+1, 1)
\STATE \textit{SVD}$_4$\texttt{<<<}grid, block\texttt{>>>}($\mathbf{A}$, $\mathbf{E}$, $\mathbf{E_r}$, $\epsilon$)
\vspace{1mm}
\ENSURE Singular values in $\mathbf{A}$ are sorted and columns of $\mathbf{E}$ and $\mathbf{E}_r$ are adjusted.
\STATE \textit{SVD}$_5$\texttt{<<<}grid, block\texttt{>>>}($\mathbf{A}$, $\mathbf{E}$, $\mathbf{E_r}$)
\vspace{1mm}
\ENSURE $\mathbf{E}$ and $\mathbf{E}_r$ are orthogonalized.
\STATE \textit{SVD}$_6$\texttt{<<<}grid, block\texttt{>>>}($\mathbf{A}$, $\mathbf{E}$, $\mathbf{E_r}$)

\end{algorithmic}
\end{algorithm}

Lines 30-32 perform QR iterations to calculate the singular value and vector matrices. Due to complex data dependencies and memory access patterns, further parallelization could introduce synchronization overhead and memory conflicts. Therefore, \textit{SVD}$_4$ is not further split, and parallelism is kept at the frequency bin ($\omega$) dimension.

Line 33 sorts the singular values in $\mathbf{A}$ and adjusts the columns of $\mathbf{E}$ and $\mathbf{E_r}$ accordingly, which is limited to parallelism at the frequency bin ($\omega$) dimension.

Finally, line 34 orthogonalizes $\mathbf{E}$ and $\mathbf{E_r}$ after column adjustment, with parallelism maintained at the frequency bin ($\omega$) dimension for the same reason as \textit{SVD}$_4$.

\section{Evaluations}
\label{sec:5}
This section presents the evaluations conducted to assess the efficiency and accuracy of our proposed implementation, with a focus on its computational performance, energy consumption and calculation precision.

\subsection{Metrics and Measurement Methods for Efficiency}
To evaluate the performance of the GPU-based implementation, we measured the processing times for both the CPU-based and GPU-based implementations, and compared the results. This analysis is crucial in determining whether the proposed GPU-based implementation can meet real-time processing requirements, and also in assessing its potential to support real-time ML/DL tasks when used as a preprocessing step in HARK. 
The comparison provides insight into the relative performance gains achieved by the proposed implementation, particularly for large-scale multi-channel audio data.


\subsection{Metrics and Measurement Methods for Calculation Accuracy}
When transitioning computations from CPU-based to GPU-based implementations, verifying the accuracy of the results is essential to ensure consistency and reliability across different hardware platforms. Variations in the algorithms, hardware architectures, and numerical precision may introduce discrepancies in the output. To evaluate consistency, we calculate the Root Mean Squared Error (RMSE) for each power spectrum ($\mathbf{P}(\theta, \omega, f)$) and compare the sound source positions detected by the GPU-based implementations with those from the CPU-based implementation across each device and processing period. This evaluates whether the GPU-based implementation can deliver reliable and consistent results when compared to the CPU-based implementation.

\subsection{Settings}
We conducts evaluations on two devices. The first one is the Jetson AGX Orin 32GB, an embedded device with an NVIDIA GPU and ARM\textsuperscript{\textregistered} Cortex\textsuperscript{\texttrademark}-A78AE v8.2 64-bit CPUs. The second one is a high-performance server with an NVIDIA A100 80GB GPU and AMD EPYC\textsuperscript{\texttrademark} 7352 CPUs, 

For testing, audio data was collected using a 60-channel microphone array, capturing sound from two distinct sources. Both CPU and GPU processing were performed in single-precision floating-point format. The HARK configuration used the default setup, with 73 frequency bins processed. The time period in eq.~\ref{eq3} was set to 50, corresponding to 0.5 seconds of

\begin{table}[t]
\small 
\caption{The average processing time of different implementations on different devices for each second of 60-channel audio. \label{table:GPU}}
\centering
\vspace{1pt}
\scalebox{1.0}{%
\begin{tabular}{lrrr}
\toprule
\textbf{Device} & \textbf{Imps.} & \textbf{\textit{SVD}} (ms)  & \textbf{SSL Module} (ms)  \\ \midrule

\multirow{2}{*}{Jetson AGX Orin} & CPU & 395.41 & 3485.09  \\
 & GPU & 0.07 & 325.16 \\ \midrule

\multirow{2}{*}{A100 Server} & CPU & 297.16  & 2659.71 \\
 & GPU & 0.07 &  155.64 \\ \bottomrule

\end{tabular}
}
\end{table}

\begin{figure}[H]
\centering
\includegraphics[scale=0.45]{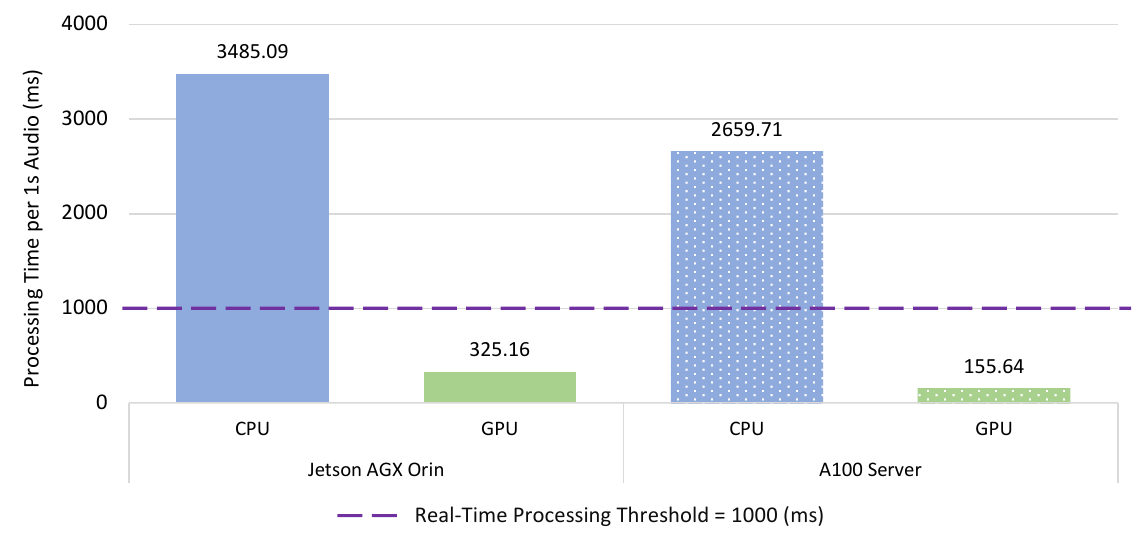}
\caption{Comparison on real-time capability for different implementations. Bars below the threshold indicate real-time capability, while bars above the threshold do not meet real-time requirements.}
\label{fig:real-time}
\end{figure}

\noindent audio. The 60-channel microphone array's transfer function includes 2522 discrete target directions, covering the full 360-degree spherical range.

\subsection{Performance Results}

Table~\ref{table:GPU} presents the performance evaluation results. The GPU-based implementation significantly outperforms the CPU-based implementation in terms of processing time. Specifically, on the Jetson AGX Orin, we observed speedups of 5648.7$\times$ for the GSVD calculation and 10.7$\times$ for the entire SSL module. On the A100 server, speedups of 4245.1$\times$ and 17.3$\times$ were observed for the GSVD calculation and the entire SSL module, respectively.

Notably, as illustrated in Fig.~\ref{fig:real-time}, the GPU-based implementation achieves real-time processing, with total processing times well below one second per second of audio data. This demonstrates the system’s capability to handle real-time SSL processing while also leaving sufficient capacity for subsequent machine learning (ML) and deep learning (DL) tasks.


\subsection{Results of Calculation Accuracy}
Table~\ref{table:Accuracy-results} shows the evaluation results for the calculation error and consistency of the sound source positions detected by the proposed implementation. The RMSE for both devices is as low as $10^{-6}$, and the proposed implementation achieved 100.00\% consistency in the detected sound source positions, confirming that the proposed implementations reliably reproduce the CPU-based SSL results.

\begin{table}[H] 
\small 
\caption{Calculation error and consistency of the sound source positions detected by the proposed implementation. \label{table:Accuracy-results}}
\vspace{1pt}
\centering
\scalebox{1.0}{%
\begin{tabular}{lrr}
\toprule
\textbf{Device} & \textbf{RMSE} & \textbf{Consistency}  \\
\midrule
Jetson AGX Orin & $2.68\times10^{-6}$ & 100.00\% \\ 
A100 Server & $2.49\times10^{-6}$ & 100.00\% \\ 

\bottomrule
\end{tabular}
}
\end{table}

\section{Conclusion}
\label{sec:6}
This paper presents GPU implementations for GSVD-based SSL, a noise-robust algorithm, with a particular focus on deploying HARK, an open-source platform for robot audition. The proposed implementation significantly reduces processing times, enabling real-time operation with a 60-channel microphone array. Specifically, on the Jetson AGX Orin, speedups of 5648.7$\times$ and 10.7$\times$ were observed for the GSVD calculation and the entire SSL module, respectively, while speedups of 4245.1$\times$ for GSVD calculation and 17.3$\times$ for the entire SSL module are obtained on a server with an NVIDIA A100 GPU and AMD EPYC\textsuperscript{\texttrademark} 7352 CPUs.

With the proposed implementation, robot audition systems can efficiently process data from large-scale microphone arrays and perform noise-robust SSL, making them more applicable for real-world applications. For instance, the GPU-based implementation allows drones to use large-scale microphone arrays for precise, noise robust and real-time SSL, enabling faster survivor detection.

However, the current implementation does not support stream processing on the 60-channel microphone array, where each audio frame is processed in real-time to ensure reliable operation in dynamic, complex environments. In the present system, audio frames are processed at fixed intervals, which introduces unavoidable delays. Future work will focus on optimizing the algorithm to improve efficiency and transitioning to a stream processing mode, enhancing both the system's responsiveness and processing accuracy.

\section*{Acknowledgement}
This work was supported by JST CREST JPMJCR19K1.

\bibliographystyle{unsrt}
\bibliography{reference} 

\vspace{12pt}

\end{document}